# Matrix and Graph Operations for Relationship Inference: An Illustration with the Kinship Inference in the China Biographical Database

[†]Chao-Lin Liu    [‡]Hongsu Wang
[†‡]Institute for Quantitative Social Science, Harvard University, USA
[†]Department of Computer Science, National Chengchi University, Taiwan
chaolin@nccu.edu.tw, hongsuwang@fas.harvard.edu

Biographical databases contain diverse information about individuals. Person names, birth information, career, friends, family and special achievements are some possible items in the record for an individual. The relationships between individuals, such as kinship and friendship, provide invaluable insights about hidden communities which are not directly recorded in databases. We show that some simple matrix and graph-based operations[1] are effective for inferring relationships among individuals, and illustrate the main ideas with the China Biographical Database[2] (CBDB).

**Relationship Matrices**

Assume that we have *n* different individuals and *m* different relationships. Let $P_i$, $i \in \Pi=\{1,...,n\}$, represent the individuals, and $R_j$, $j \in \Lambda=\{1,...,m\}$, represent the relationships. We can store the relationships between the individuals with a matrix **T**,

$$T = \begin{array}{c} \\ 1 \\ 2 \\ 3 \end{array} \begin{bmatrix} 1 & 2 & 3 \\ & F & \\ & & DHB \\ & & \end{bmatrix}$$

**Fig. 1. A relationship matrix for kinship**

in which a cell $T_{1,2}$ is the relationship between $P_1$ and $P_2$ that is recorded in a database. In CBDB, an $R_1$ can be **basic**, e.g., "F" which represents a "father-of" relationship, and an $R_2$ can be **extended**, e.g., "DHB" which represents a "daughter's husband's brother"[3] relationship while the participating individuals in between are unknown. In the matrix in Figure 1, we have $T_{1,2}$="F", and that means $P_2$ is a father of $P_1$. That $T_{2,3}$="DHB" means $P_3$ is $P_2$'s daughter's husband's brother.

A **relationship matrix** like that in Figure 1 does not have to be symmetric. In addition, a cell in the matrix may need to accommodate multiple values when necessary. When we convert the relationship matrix directly from a biographical database, the original matrix may be asymmetric, because the information stored in databases are not symmetric. If we wish, we can build a symmetric matrix from a raw matrix like that in Figure 1, by checking and combining $T_{i1,i2}$ and $T_{i2,i1}$ for any combination of $i1 \in \Pi$ and $i2 \in \Pi$. In this process, we might find conflicting or even parallel relationships between two individuals, which will need domain experts for further inspection. A person may have parallel (or multiple) relationships with another due to many reasons.

---

[1] Chao-Lin Liu, Tun-Wen Pai, Chun-Tien Chang, and Chang-Ming Hsieh (2001) Path-planning algorithms for public transportation systems, *Proc. of the Fourth Int'l IEEE Conf. on Intelligent Transportation Systems*, 1061–1066.
[2] China Biographical Database: https://projects.iq.harvard.edu/cbdb
[3] Here, D stands for "daughter", H for "husband", and B for "brother".

We can define operations for relationship inferences with a relationship matrix (RM). The simplest choice is to convert cells that have known relationships into 1s and others into 0s. We convert the matrix in Figure 1 into the **M** and compute its square in Figure 2. Note that the 1 in $\mathbf{M}^2$ is informative. It indicates that $P_3$ is $P_1$'s father's daughter's husband's brother.

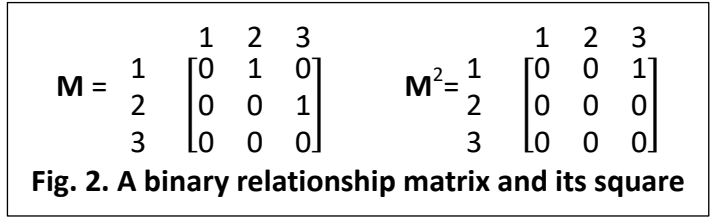

Fig. 2. A binary relationship matrix and its square

The fact that the powers of an RM are informative is a natural result of the definition of multiplication of matrices. Let $\mathbf{M}^2_{x,y}$ denote the cell (x,y) of $\mathbf{M}^2$. We compute $\mathbf{M}^2_{x,y}$ based on the definition in (1). When we convert an RM into a binary RM, a cell is one only if the cell represents an existing relationship. Hence, definition (1) essentially states that $P_x$ and $P_y$ has relationships if there exists a person $P_z$ such that (1) $P_z$ is a relative of $P_x$ and (2) $P_y$ is a relative of $P_z$. Furthermore, the value of $\mathbf{M}^2_{x,y}$ is the total number of such two-step relationships.

$$\mathbf{M}^2_{x,y} = \sum_{z\in \Pi} \mathbf{M}_{x,z} \times \mathbf{M}_{z,y} \qquad (1)$$

In addition to counting the indirect relationships when computing (1), we can concatenate the relationships of $\mathbf{M}_{x,z}$ and $\mathbf{M}_{z,y}$ to obtain a two-step relationship for $P_x$ and $P_y$. For instance, concatenating "F", "2", and "DHB" to obtain "1_F_2_DHB_3", in the computation for the $\mathbf{M}^2$ in Figure 2, will record one indirect relationship between $P_1$ and $P_3$ and the intermediate participants. This operation of concatenation and recording can be achieved by software easily.

**Applications and Extensions**

We can easily generalize the implication of (1) to consider the semantics of the values in the n-th power of a binary RM, **M**, to achieve the following theorem.

**Theorem 1**. Let **M** denote a binary relationship matrix as we explained above. A person $P_x$ and a person $P_y$ are relatives if there is a $\rho \geq 1$ such that $\mathbf{M}^\rho_{x,y} > 0$.

There are 360,000 individuals in the 2015 version of CBDB, and we should be able to identify clusters of individuals that represent families. Although Theorem 1 is applicable for this task, a more efficient method is to apply the ideas of identifying connected components in mathematical graphs. The basic idea is quite simple: individuals that have relationships belong to a family and members of different families cannot have any relationships.

Corollary 1 allows us to find the shortest relationships between two individuals. If the kin relationships in the original RM are genuinely basic, e.g., father, month, son, daughter, husband, wife, brother, and sister, then we could find the most direct relationships between individuals with the corollary.

**Corollary 1**. Let **M** denote the binary RM of an RM. By finding the smallest $\sigma$ such that $\mathbf{M}^\sigma_{x,y}$ is positive, we identify the relationship between $P_x$ and $P_y$ that has the fewest number of intermediate participants.

In practice, not all kin relationships recorded in historical documents are basic, and extended relationships such as "DHB" may be recorded. As a consequence, a "shortest path" as defined in Corollary 1 may not be the most direct path between two individuals.

We can define the "**length**s" of relationships to quantify the distances between individuals.[4] Take "DHB" as an example. That $P_y$ is a "DHB" of $P_x$ means that $P_y$ is one generation after $P_x$, and we can define the **g-len** of "DHB" as -1[5]. Analogously, the g-len of "F" is 1. The g-len of a path is the sum of the g-lens of all relationships in the path, so the g-len of "F_2_DHB" is 0. The length of a relationship can encode the "sideway" distances as well. We can define the **s-len**s of "DHB" and "F" to be 2 and 0, respectively. In practice, there are only few hundreds of relationships in CBDB, so it is feasible to annotate the relationships with their lengths.

**Building Family Networks**

The upper part of Figure 3 shows three actual paths[6] between individuals in CBDB. We can construct a family network that is **consistent** with the paths, and provide the network to domain experts for verification.

NAME3_B_NAME4_S_NAME5 (i.e., NAME3's brother is NAME4; NAME4's son is NAME5)
NAME3_FF_NAME1_SS_NAME4_S_NAME5
NAME3_F_NAME2_S_NAME4_S_NAME5

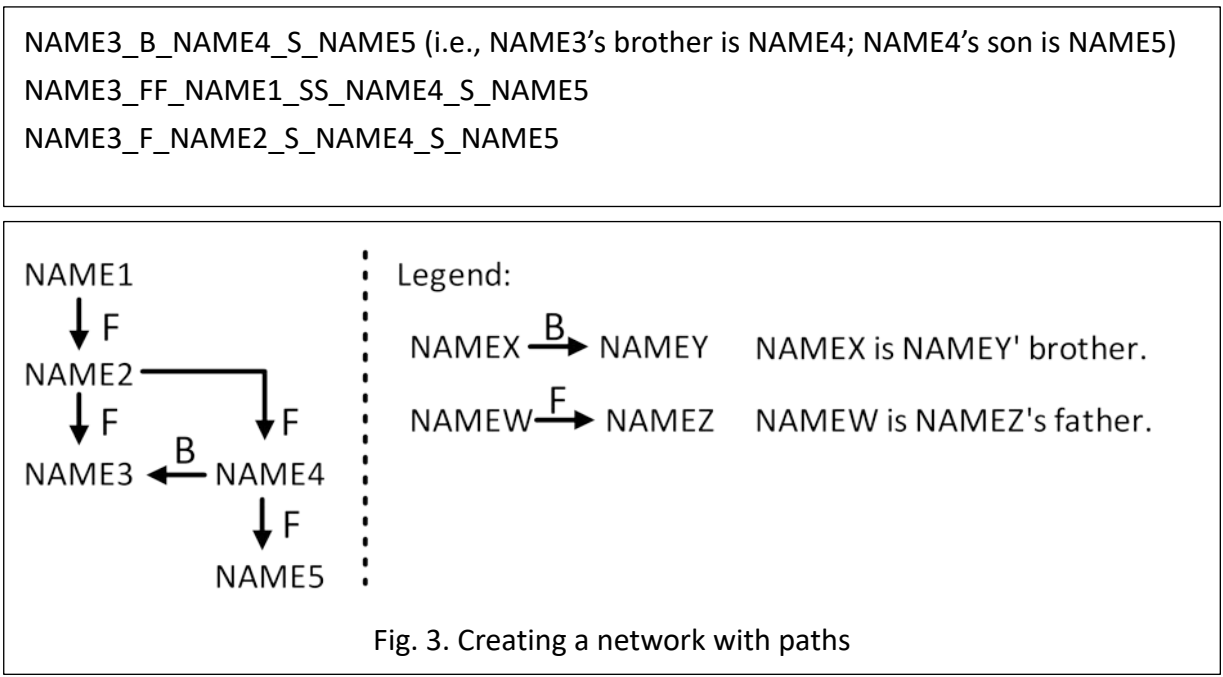

Fig. 3. Creating a network with paths

---

[4] Similar quantification methods were adopted by Professor Michael Fuller when he designed a backend service for CBDB. Similar quantification methods were also described in the following working paper.
  Ke Deng, Peter K. Bol, Stuart M. Shieber, and Jun S. Liu. Building a Kinship Network for Political Figures in History.
[5] D is a relationship from an earlier generation to a later generation (-1). Both H and B are relations for the same generations (+0). Hence the g-cost of "DHB" is -1+0+0=-1.
[6] The actual names in Chinese are NAME1 for 呂弸中 (lu3 peng2 zhong1), NAME2 for 呂大器 (lu3 da4 qi4), NAME3 for 呂祖謙 (lu3 zu3 qian1), NAME4 for 呂祖儉 (lu3 zu3 jian3), and NAME5 for 呂喬年 (lu3 qiao2 nian3).